\documentclass[mathleft]{an}
\usepackage{graphicx}
\usepackage{times}
\begin{document}

\Pagespan{789}{}
\Yearpublication{2010}%
\Yearsubmission{2010}%
\Month{11}%
\Volume{999}%
\Issue{88}%

\title{The variability of the CoRoT target HD171834: 
$\gamma$\,Dor pulsations and/or activity?\,\thanks{Based on observations made with ESO Telescopes at the La Silla Observatory under  the ESO Large Programmes ESO LP 178.D-0361 and ESO LP 182.D-0356, and on data collected at the Centro Astron\'omico Hispano Alem\'an (CAHA) at Calar Alto, operated jointly by the Max-Planck Institut f\"ur Astronomie and the Instituto de Astrof\'{\i}sica de Andaluc\'{\i}a (CSIC). Based on observations made with the Nordic Optical Telescope, operated on the island of La Palma jointly by Denmark, Finland, Iceland,
Norway, and Sweden, in the Spanish Observatorio del Roque de los
Muchachos of the Instituto de Astrof\'{\i}sica de Canarias.  Also based on observations made at Observatoire de Haute Provence (France) and at  Mount John University Observatory (New Zealand). The CoRoT space mission has been developed and
is operated by the French Space agency CNES in collaboration with the
Science
Programs of ESA, ESA, Austria, Belgium, Brazil, Germany and
Spain.}}

\author{K. Uytterhoeven\inst{1}\fnmsep\thanks{\email{katrien.uytterhoeven@cea.fr}}, P. Mathias\inst{2}, A. Baglin\inst{3}, M. Rainer\inst{4}, E. Poretti\inst{4},  P. Amado\inst{5}, E. Chapellier\inst{6}, L. Mantegazza\inst{4}, K. Pollard\inst{7}, J.C. Suarez\inst{5}, P.M. Kilmartin\inst{7}, K.H. Sato\inst{1},  R.A. Garc\'{\i}a\inst{1}, M. Auvergne\inst{3}, E. Michel\inst{3}, R. Samadi\inst{3}, C. Catala\inst{3}, \and F. Baudin\inst{8}
}
\titlerunning{The variability of the CoRoT target HD171834}
\authorrunning{K. Uytterhoeven et al.}
\institute{
Laboratoire AIM, CEA/DSM-CNRS-Universit\'e Paris Diderot; CEA, IRFU, SAp, centre de Saclay, 91191, Gif-sur-Yvette, France
\and 
Laboratoire d'Astrophysique de Toulouse-Tarbes, Universit\'e de Toulouse, CNRS, 57 Avenue Azereix, 65000 Tarbes, France
\and 
LESIA, UMR8109, Universit\'e Pierre et Marie Curie, Universit\'e Denis Diderot, Observatoire de Paris, 92195 Meudon, France
\and
INAF-Osservatorio Astronomico di Brera, Via E. Bianchi 46, 23807 Merate, Italy
\and 
Instituto de Astrof\'{\i}sica de Andaluc\'{\i}a (CSIC), Apartado 3004, 18080 Granada, Spain
\and 
UMR 6525 H. Fizeau, UNS, CNRS, OCA, Campus Valrose, 06108 Nice Cedex 2, France
\and
Dep.\, of Physics and Astronomy, University of Canterbury, Private Bag 4800, Christchurch, New Zealand
\and
Institut d'Astrophysique Spatiale, UMR 8617, Universit\'e Paris XI, B\^atiment 121, 91405 Orsay Cedex, France
}

\received{01 April 2010}
\accepted{--}
\publonline{later}

\keywords{stars: oscillations, stars: individual (HD171834), stars: variables: gamma Doradus stars}

\abstract{We present the preliminary results of a frequency and line-profile analysis of the CoRoT $\gamma$\,Dor candidate HD171834. The data consist of 149 days of CoRoT light curves and a ground-based dataset of more than 1400 high-resolution spectra, obtained with six different instruments. Low-amplitude frequencies between 0 and 5\,d$^{-1}$, dominated by a frequency near 0.96\,d$^{-1}$ and several of its harmonics, are detected. These findings suggest that HD171834 is not a mere $\gamma$\,Dor pulsator and that stellar activity plays an important role in its variable behaviour.}

\maketitle

\section{Introduction}

It are exciting  times for seismic studies of $\gamma$\,Dor pulsators, thanks to the successful operation of asteroseismic space missions, such as CoRoT  (Baglin et al. 2006) and Kepler  (Gilliland et al. 2010). The class of $\gamma$\,Dor stars (Kaye et al. 1999) consists of stars of spectral types A-F that are slightly more massive than the Sun (1.2Msun $<$ M $<$ 2.5Msun).  They pulsate in high-order, non-radial gravity (g-) mode pulsations, explained in terms of a flux modulation induced by the upper convective layer  (Guzik et
al. 2000; Dupret et al. 2005; Grigahc\`ene 2005).  As only g-modes allow the probing of the deep stellar interior, $\gamma$\,Dor stars are  extremely interesting asteroseismic targets (e.g. Miglio et al. 2008). 

The $\gamma$\,Dor pulsators are however very challenging targets, both from an observational as theoretical point of view. Their pulsation periods are of order of a day and hence very difficult to monitor from the ground. Moreover, the corresponding pulsation amplitudes are fairly small (below 0.05mag and 2 km s$^{-1}$). Several ground-based observational efforts have been undertaken  to describe the pulsational behaviour in as many class members as possible (e.g. Poretti et al. 2002; Mathias et al. 2004; De Cat et al. 2006; Rodr\'{\i}guez et al. 2006a,b;  Uytterhoeven et al. 2008; Cuypers et al. 2009), resulting in the detection and identification of only a limited amount of frequencies and associated mode parameters.  Theoretically, the fast rotation observed in several $\gamma$\,Dor stars is posing problems in the description of the pulsational instability, as the current models do not account for higher-order rotational effects (Su\'arez et al. 2005; Bouabid et al. 2008, 2009; Moya et al. 2008). 

The CoRoT satellite disclosed for the first time the very rich and complicated frequency spectrum, consisting of several hundreds of frequencies, of a $\gamma$\,Dor star (HD\,49434, Chapellier et al. 2010). The first Kepler data of hundreds of $\gamma$\,Dor candidates reveal similar dense frequency spectra (Grigahc\`ene et al. 2010), opening new prospectives in the study of the so far not-well understood class of $\gamma$\,Dor pulsators. Moreover, the space data promise further investigation of the nature of hybrid $\gamma$\,Dor-$\delta$\,Sct pulsators, i.e. stars that pulsate in $g-$ and $p-$modes simultaneously, as several hybrid candidates have been identified among the CoRoT and Kepler targets (Mathias et al. 2009; Grigahc\`ene et al. 2010). 

\section{The $\gamma$ Dor candidate HD171834}
The target of this paper is HD171834 (HIP91237; F3V; Vmag = 5.44). The star lies close to the red border of the  $\gamma$\,Dor instability strip (see Fig.1), with effective temperature  $T_{\rm eff}$ = 6750$\pm$250 K, surface gravity  $\log g$ = 3.9$\pm$0.25, and metallicity [Fe/H]=-0.5$\pm$0.15 (Lastenet et al. 2001).  The $\gamma$\,Dor candidate HD171834 appears to be photometrically constant from the ground (Poretti et al. 2003), but weak line-profile variations with low amplitudes have been reported by Mathias et al. (2004). 

\begin{figure}
\centering
\rotatebox{-90}{\includegraphics[width=0.30\textwidth]{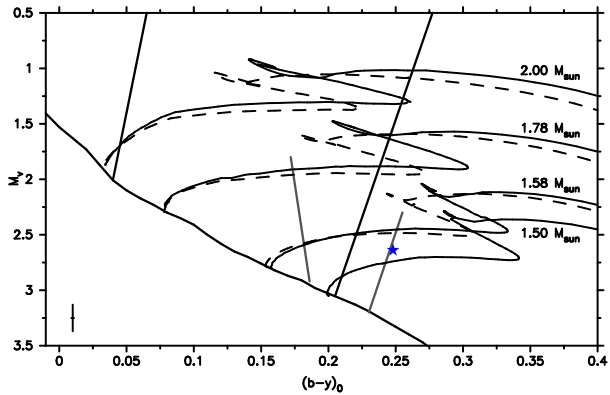}}
\caption{Colour-magnitude diagram with the position of HD171834 indicated (star). Dashed and solid tracks are evolutionary tracks for the overshooting extension distances 0.1 and 0.2 (Claret 1995). The ZAMS, the borders of the $\delta$\,Sct (black) and $\gamma$\,Dor (gray) instability strips are indicated by full lines (Dupret et al. 2005).}
\label{CMD}
\end{figure}

\section{The CoRoT  time-series}
\begin{figure}
\centering
\includegraphics[width=0.45\textwidth]{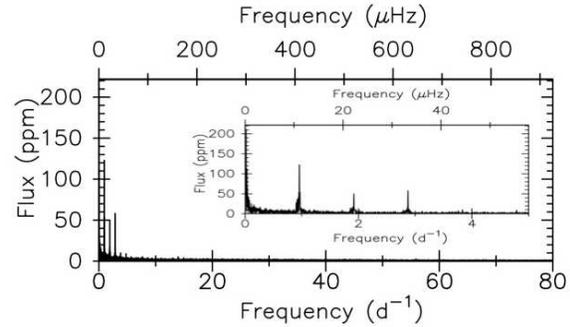}
\caption{Amplitude spectrum of the CoRoT data of HD171834. The inset shows the frequency region 0-5d$^{-1}$ in more detail.}
\label{Scargle}
\end{figure}

The CoRoT satellite observed HD171834 for 149 days during its second Long Run in the center direction (LRc2, April-September 2008). To optimise the quality of the light curves we corrected for time gaps caused by the passage of the satellite through the South Atlantic Anomaly using the inpainting algorithm  (Sato et al. 2010).   The inpainted light curves were subsequently detrended and analysed in frequency with the Van\'{\i}\v{c}ek method (Van\'{\i}\v{c}ek 1971). The amplitude periodogram shows no obvious peaks for frequencies higher than 15\,d$^{-1}$ (see Fig.~\ref{Scargle}). The highest amplitudes are reached for frequencies between 0 and 5\,d$^{-1}$, with 0.960\,d$^{-1}$ as dominant frequency. Harmonic frequencies seem to play an important role.

\section{Ground-based spectroscopic time-series}
In the framework of the ground-based follow-up observations of CoRoT targets (Uytterhoeven et al. 2008a, 2009; Uytterhoeven 2009) HD171834 was monitored for two seasons (2008 and  2009) with six different instruments (see Table\ref{logbook}) at the Observatory of Calar Alto (CaHa, Spain), European Southern Observatory (ESO, Chile), Observatoire de Haute Provence (OHP, France), Mount John University Observatory (MJUO, New Zealand), and Observatorio Roque de los Muchachos, La Palma (ORM, Spain). Never has there been a time-series with more than 1400 spectra for any other $\gamma$ Dor star. The analysis was carried out on the Least-Squares Deconvolved (LSD, Donati et al. 1997) spectra, calculated with mask $T_{\rm eff}=6750$K and $\log g$=4.0, that were subsequently normalised and corrected for instrumental radial velocity shifts (see Uytterhoeven et al. 2008 for a description of the process).   Figure~\ref{mom1} shows the radial velocities ($\langle v\rangle$) calculated from the LSD profiles. We did not find evidence for a possibile binary nature of HD171834 in the spectra.

\begin{table*}
 \begin{center}
\caption{Logbook of the observations of HD171834 in 2008 and 2009. The columns give information on  the instrument and observatory, the number of spectra (N), the timespan ($\Delta$T, in days) of the dataset, the resolution of the spectrograph, and the observers.}
\label{logbook}
\begin{tabular}{lcccccc} \hline
Instrument & Observatory&N & $\Delta$T  & $\Delta$T & Resolution & observer\\
           &    & & 2008       & 2009      &            &  \\ \hline 
FOCES@2.2m& CaHa &562 & 25.2 & 34.1 & 40,000 & PJA\\
FEROS@2.2m & ESO& 193 & 25.2 & --   & 48,000 & LM, KU\\
SOPHIE@1.93m & OHP& 471 & 29.2 & 31.1 & 40,000 & PM, KU\\
HERCULES@1.0m & MJUO&55 & 9.1 & -- & 35,000 & KP, PMK\\
FIES@NOT & ORM& 46 & -- & 3.4&  46,000 & KU\\
HARPS@3.6m &ESO& 79 & -- & 39.1 & 80,000 & EP, JCS\\ \hline
\end{tabular}
\end{center}
\end{table*}

\begin{figure}
\centering
\rotatebox{-90}{\includegraphics[width=0.45\textwidth]{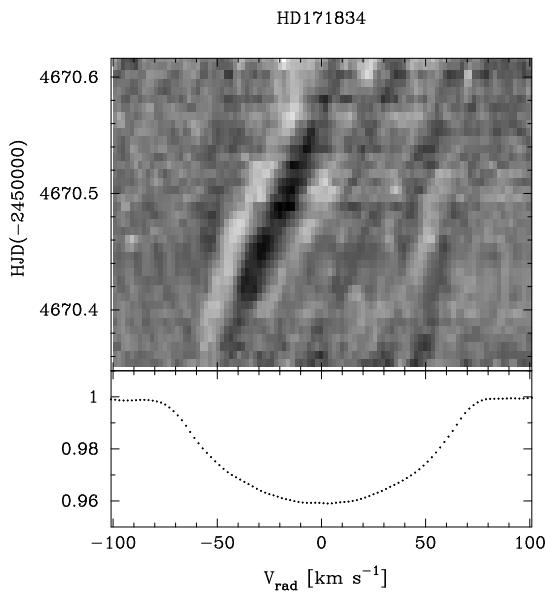}}
\caption{Grayscale representation of one night of SOPHIE observations (HJD 2454670) illustrating the typical nightly  line-profile variable behaviour of HD171834. Bottom: Average LSD profile calculated from the SOPHIE time-series consisting of 471 spectra. Top: Residual LSD profiles (HJD 2454670) with respect to the average LSD profile.}
\label{grayscale}
\end{figure}

\subsection{Line-profile analysis}
\begin{figure}
\centering
\rotatebox{-90}{\includegraphics[width=0.30\textwidth]{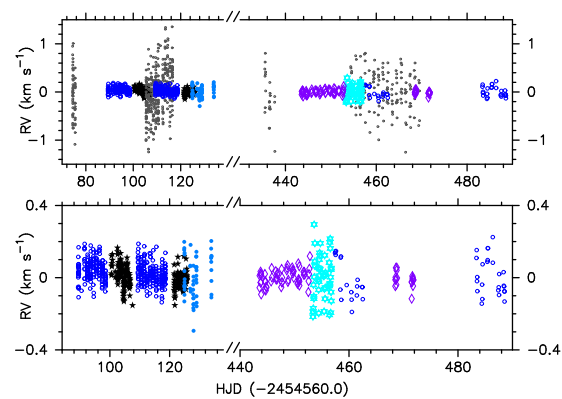}}
\caption{Radial velocities ($\langle v\rangle$) calculated from the LSD profiles of FEROS (black filled stars), SOPHIE (blue open circles), FOCES (gray circles), HERCULES (blue filled circles), HARPS (purple diamonds), and FIES (cyan open stars) spectra (top) obtained in 2008 and 2009. In the bottom panel we did not plot the more dispersed FOCES spectra, illustrating the low amplitude variability of HD171834.}
\label{mom1}
\end{figure}

The typical nightly line-profile variable behaviour of HD171834 is illustrated in Fig.~\ref{grayscale}. Narrow, well-defined line-profile  \textquoteleft\,bumps\,\textquoteright \newline are seen to propagate through the line-profile with periods longer than a day. The individual datasets are too short to reveal  significant long-term variabilities. Therefore, we analysed the combined datasets, spanning 414 days, and excluded the more dispersed FOCES spectra as they introduced spurious frequencies (see bottom panel Fig.~\ref{mom1}). To search for periodicities in the time-series of LSD spectra we used the Van\'{\i}\v{c}ek method on the velocity moments (Aerts et al. 1992), and performed a pixel-to-pixel analysis using the Intensity Period Search method (IPS, Telting \& Schrijvers 1997). In the variations of the first moment the CoRoT frequency near 0.96\,d$^{-1}$ is recovered, together with several of its harmonics.  The IPS method shows several low-amplitude frequencies between 0.1 and 1.0\,d$^{-1}$, and harmonics of 0.96\,d$^{-1}$.

\section{Conclusions}
 CoRoT data reveal that the $\gamma$ Dor candidate HD171834 shows low-amplitude frequencies between 0 and 5\,d$^{-1}$, dominated by 0.96\,d$^{-1}$ and several of its harmonics, suggesting stellar activity. The long-period low-amplitude variations are difficult to detect from the ground. Only by combining all available spectra, spanning more than one season and consisting of more than 1400 spectra are the frequencies  recovered. A more detailed analysis is ongoing to confirm the double nature of the variability, rotational and pulsational, and their possible connection  and interplay. Also, spectropolarimetric observations  are scheduled with NARVAL@TBL to investigate the possible importance of a magnetic field. 

\begin{acknowledgements}
MR, EP, and LM acknowledge financial support from  the Italian
ESS project, contract ASI/INAF I/015/07/0, WP 03170.
EP acknowledges the support from the European Helio and Asteroseismology
Network (HELAS) for the participation to the Conference.
This work has been partially funded by the GOLF/CNES grant at the CEA/Saclay.
\end{acknowledgements}


\begin{thebibliography}{}
\bibitem{} Aerts, C., De Pauw, M.,  Waelkens, C.: 1992, A\&A 266, 294
\bibitem{} Baglin, A., Auvergne, M., Boisnard, L,. et al.: 2006, in 36th COSPAR Scientific Assembly 36, 3749
\bibitem[]{} Bouabid, M.-P, Uytterhoeven, K., Miglio, A., et al.: 2008, CoAst 157, 290
\bibitem[Bouabid et al.(2009)]{Boua09}
   Bouabid, M.~-P., Montalb{\'a}n, J., Miglio, A., Dupret, M.~-A., Grigahc{\`e}ne, A.,  Noels, A.: 2009,  AIPC 1170, 477
\bibitem[Chapellier et al. (2010)]{CoRoTHD49434} Chapellier, E., Rod\'{\i}guez, E., Auvergne, M., et al.: 2010, A\&A, submitted
\bibitem[Claret (1995)]{Claret} Claret, A.: 1995, A\&ASS 109, 441
\bibitem{} Cuypers, J., Aerts, C., De Cat, P., et al.: 2009, A\&A 499, 967
 \bibitem[De Cat et al. (2006)]{Peter06} De Cat, P., Eyer, L., Cuypers, J., et al.: 2006, A\&A 449, 281
\bibitem[Donati et al. (1997)]{d97}
Donati, J.-F., Semel, M., Carter, B.D., Rees, D. E.,  Collier Cameron, A.: 1997, MNRAS 291, 658
\bibitem[Dupret et al.(2005)] {Dupt05} 
   Dupret, M.~-A., Grigahc{\`e}ne, A., Garrido, R., Gabriel, M.,  Scuflaire, R.: 2005, A\&A 435, 927
\bibitem{} Gilliland, R.L., Brown, T.M., Christensen-Dalsgaard, J., et al.: 2010, PASP 122, 131
\bibitem[Grigahc{\`e}ne et al.(2005)] {Grig05} 
   Grigahc{\`e}ne, A., Dupret, M.~-A., Gabriel, M., Garrido, R.,  Scuflaire, R.: 2005, A\&A 434, 1055
\bibitem[Grigahc{\`e}ne et al.(2010)] {Grig10} 
   Grigahc{\`e}ne, A., Antoci, V., Balona, L., et al.: 2010, ApJ 713, 192
\bibitem[Guzik et al.(2000)]{Guzik00} 
   Guzik, J.~A., Kaye, A.~B., Bradley, P.~A., Cox, A.~N.,  Neuforge, C.: 2000, \apj 542, 57
\bibitem[]{} Kaye, A.B., Handler, G., Krisciunas, K., Poretti, E.,  Zerbi, F.M.: 1999, PASP 111, 840
\bibitem[Lastenet et al. (2001)]{Lastenet} Lastennet et al.: 2001, A\&A 365, 535
\bibitem[Mathias et al. (2004)]{Math04} Mathias et al.: 2004, A\&A 417, 189
\bibitem[Mathias et al.(2009)] {Math09}    Mathias, P., Chapellier, E., Bouabid, M., et al.: 2009, AIPC 1170, 486
\bibitem{}  Moya, A., Christensen-Dalsgaard, J., Charpinet, S., et al.: 2008, AP\&SS 316, 231
\bibitem[]{} Miglio, A., Montalb\'an, J., Noels,  A., Eggenberger, P.: 2008, MNRAS 386, 1487
\bibitem{} Poretti, E., Koen, C., Bossi, M., et al.: 2002, A\&A 384, 513
\bibitem[Poretti et al. (2003)]{Poretti03} Poretti et al.: 2003, A\&A 406, 203
\bibitem{} Rodr\'{\i}guez, E., Amado, P.J.,  Su\'arez, J.C.,, et al.: 2006a, A\&A 450, 715
\bibitem{} Rodr\'{\i}guez, E., Costa, V., Zhou, A.-Y., et al.: 2006b, A\&A 456, 261
\bibitem[Sato et al. (2010)]{Sato} Sato, K.H., Garc\'{\i}a, R.A., Pires, S., et al.,: 2010, AN, submitted (this volume) (arXiv:1003.5178)
\bibitem{} Su\'arez, J.C., Moya, A., Mart\'{\i}n-Ru\'{\i}z, S., Amado, P.J., Grigahc\`ene, A., Garrido, R.: 2005, A\&A 443, 271 
\bibitem[]{}  Telting, J.H.,  Schrijvers, C.: 1997, A\&A 317, 723
\bibitem[Uytterhoeven et al. (2008a)]{CoRoTproc08} Uytterhoeven, K., Poretti, E., Rainer, M., et al.: 2008a, Journal of Physics Conf. Ser. 118, 2077
\bibitem[Uytterhoeven et al. (2008b)]{HD49434} Uytterhoeven, K., Mathias, P., Poretti, E., et al.: 2008b, A\&A 489, 2213
 \bibitem[Uytterhoeven (2009)]{procLiege} Uytterhoeven, K.: 2009, CoAst 158, 156 
\bibitem[Uytterhoeven et al. (2009)]{procSantaFe} Uytterhoeven, K., Poretti, E., Mathias, P., et al.: 2009, AIP Conf. Proc. 1170, 327
\bibitem[Van\'{\i}cek (1971)]{vani} Van\'{\i}\v{c}ek, P.: 1971, Ap\&SS 12, 10
\end{thebibliography}
\end{document}